# Detection of Missing Assemblies and Estimation of the Scattering Densities in a VSC-24 Dry Storage Cask with Cosmic-Ray-Muon-Based Computed Tomography

Zhengzhi Liu[1], Jason Hayward[1], Can Liao[2] and Haori Yang[2]

[1]University of Tennessee-Knoxville, TN 37996, USA

[2]Oregon State University, Corvallis, OR 97331, USA

**Abstract:** Highly energetic, cosmic-ray muons can penetrate a dry storage cask and yield information about the material inside it by making use of the physics of multiple Coulomb scattering. Work by others has shown this information may be used for verification of dry storage cask contents after continuity of knowledge has been lost. In our modeling and simulation approach, we use ideal planar radiation detectors to record the trajectories and momentum of both incident and exiting cosmic ray muons; this choice allows us to demonstrate the fundamental limit of the technology for a particular measurement and reconstruction method. In a method analogous to computed tomography with the attenuation coefficient replaced by scattering density, we apply a filtered back projection algorithm in order to reconstruct the geometry in modeled scenarios for a VSC-24 concrete-walled cask. We also report on our attempt to estimate material-specific information. A scenario where one of the middle four spent nuclear fuel assemblies is missing—undetectable with a simple PoCA-based approach—is expected to be detectable with a CT-based approach. Moreover, a trickier scenario where one or more assemblies is replaced by a dummy assembly is put forward. In this case, we expect that this dry storage cask should be found to be not as declared based on our simulation and reconstruction results.

Keywords: Cosmic ray muon, dry storage cask, scattering density, computed tomography, VSC-24

**Introduction**

Given the abandonment of commercial reprocessing in the United States [1] and the failure to open a permanent geologic repository for storage of spent nuclear fuel [2], alternative storage solutions are needed in the interim. Offloading of older, cooler spent fuel into concrete dry storage casks (DSC) [3] that are stored on-site is the expedient solution. Due to nuclear proliferation concerns and the high expense of resealing storage casks, it is pressing to develop a nondestructive monitoring system to verify the contents of a cask once continuity of knowledge has been lost. In this paper, we address imaging a dry storage cask with cosmic ray muons based on a computed tomography technique [4].

The muon is an elementary particle similar to electron, with a charge of +1e or -1e and a spin of $\frac{1}{2}$, but with a much greater mass (~207 $m_e$). Muons are created when primary cosmic rays, primarily protons, collide with molecules in the earth's upper atmosphere [5]. These naturally occurring particles are the dominant component of cosmic radiation flux in the atmosphere. The flux is approximately 10,000 muons/m²/min at sea level, dropping off roughly as $cos^2\theta_z$, where $\theta_z$ is the zenith angle. The range of muon energies is wide, ranging from about 100 MeV to 10 GeV, with the average value being 3-4 GeV [6]. Both the flux and energy vary with a number of factors, including polar angle, elevation, and the solar cycle. Cosmic ray muons interact with matter in two primary ways: electromagnetic interactions with electrons including ionization and excitation; and multiple Coulomb scattering from nuclei [7]. Compared with ionization and excitation interactions, multiple Coulomb scattering is more sensitive to the atomic number of the material [8] [9].

In mathematics, the Radon transform is the integral transform which takes a function $f$ defined on a plane to a function $R_f$ defined on the (two-dimensional) space of lines in a plane, whose value at a particular line is equal to the line integral of the function over that line.

$$R_f(L) = \int_L f(\boldsymbol{X})|d\boldsymbol{X}| \tag{1}$$

The transform was introduced in 1917 by Johann Radon [10], who also provided a formula for the inverse transform. Radon further included formulas for the transform in three dimensions, in which the integral is taken over planes. Development of computer assisted tomography based on this theorem to see the human body via X-ray images was made by Allan M. Cormack [11] and Godfrey N. Hounsfield [12], for which both won the Nobel Prize in Physiology or Medicine in 1979.

Previous work in using muons to image objects either used a simple PoCA method [13], [14], which is fundamentally incapable of resolving the fine structure of an imaged object, or statistical reconstruction[15], which is extremely computationally expensive. The latest work of imaging a dry storage cask with computed tomography is described in [16]; in this work, the authors use only the horizontal directions of the incident muon tracks to determine the position bin where the muon scattering angle is stored. Without measuring the muon momentum, CT reconstruction is carried out with a multigroup model [17], [18] to infer the geometrical layout in a DSC. In this work, which does not focus on what can be achieved with specific detector technologies, we assume that the momentum of muons is perfectly measurable using ideal planar detectors. This assumption has been made in other related work, e.g., [15], [19]. To date, there have been a couple of methods developed to measure muon momentum. A typical spectrometer envisaged for the LHC can achieve 10% energy resolution limited for low energy muons[20] and ATLAS detector can yield a relative resolution better than 3% over a wide $P_T$ range[21]. Even though indiscriminately deeming naturally existing muons as monoenergetic, with the same method, the reconstructed image is still clear enough to tell whether there is a spent fuel bundle missing in DSC, but material information will be lost. Instead of using a multigroup muon model, we use each muon's momentum which is precisely measured to correct for the influence of polyenergetic muons. Both the horizontal directions of the incident muons and their PoCA points are used to project the scattering angles toward corresponding bins. Next, we apply filtered back projection to the collected information (sinogram) in order to calculate the scattering density of each pixel.

Due to the inaccuracy of the PoCA assumption, it's impossible with that method to identify one missing spent nuclear fuel assembly in a cask, especially the middle one[14]. We examine a different method which has the potential of greatly improved results. In this paper, we focus on difficult cases including a missing middle assembly or the replacement of a middle assembly. Seven different situations are simulated: one of the middle four assemblies missing; the middle four assemblies are replaced by copper assemblies, by lead assemblies or by tungsten assemblies; one of the middle four assemblies is replaced by copper assembly, by a lead assembly and by a tungsten assembly.

## 2D muon CT theory

In transmission-based medical imaging, the incident beam is usually made of x-rays, which, in contrast to muons, are neutrally charged particles. This beam typically undergoes the photoelectric effect, Compton scattering, pair production (if E > 1.022 MeV), or no reaction, as it traverses through a patient or, most generally, an object. The incident beam often has a significant probability of experiencing Compton scattering in an object, which can scatter x-rays at large angles. Thus, the detected beam flux is typically the uncollided beam. As illustrated in Figure 1, let $I_0$ and $I$ be the incident beam and outgoing beam intensities, respectively. The ratio $I/I_0$ is often used to reconstruct the investigated object using filtered back projection [22]. Of course, the signal obtained from one projection or view is not enough to reconstruct an image. Thus, one typically rotates the radiation source and detectors together, while the object remains fixed, in order to obtain additional information from other views.

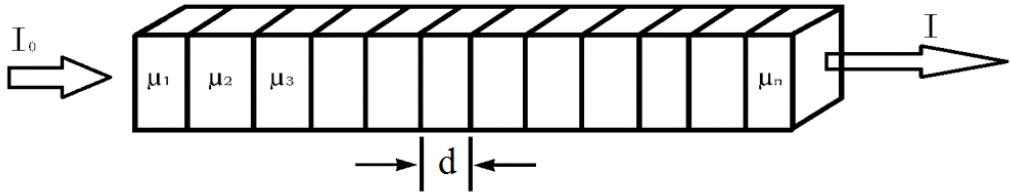

Figure 1. Illustration of a neutral beam crossing a discretized object.

Referring to *Figure 1*, the intensity can be described by

$$I = I_0 e^{-d \sum_{i=1}^{n} \mu_i} \qquad (2)$$

where $d$ is a selected discretized length in cm and $\mu_i$ is the attenuation coefficient of the $i^{th}$ pixel in cm$^{-1}$. After rearrangement,

$$\ln\left(\frac{I_0}{I}\right) = d \sum_{i=1}^{n} \mu_i \qquad (3)$$

In our application of imaging a dry storage cask containing spent nuclear fuel, the incident source is naturally occurring cosmic ray muons. Most muons are transmitted through objects [23], especially in the case of muons with high momentum (compared with the mean that falls in the range 3-4 GeV). Even though the transmission ratio $I_0/I$ could be used to reconstruct the inner configuration of a DSC, it is not likely to yield information about the specific materials through which the muon passes.

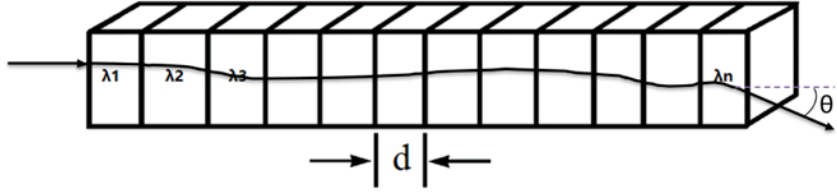

Figure 2. Illustration of a muon traversing a discretized object. The magnitude of the scattering angle is exaggerated in the figure for the purpose of illustration.

Ionization leads to energy loss of muons, and multiple Coulomb scattering causes muons to deviate from a straight line path, as illustrated in Figure 2. When many muons traverse an object, many different scattering angles will be registered, following a Gaussian distribution with zero mean value and a standard deviation $\sigma_\theta$ [24] given by

$$\sigma_\theta \cong \frac{15 MeV}{\beta cp} \sqrt{\frac{L}{L_{rad}}} \qquad (4)$$

where $p$ is the muon's momentum in MeV/c, $L$ is the length of the object, and $L_{rad}$ is the radiation length of the material. For the $i^{th}$ voxel, the variance is given by

$$\sigma_{\theta_i}^2 = d\lambda_i \qquad (5)$$

where $\lambda_i$ is the scattering density of the $i^{th}$ pixel. Since the multiple Coulomb scattering in individual pixels can be treated as independent, the variance of the ray signal may be written as

$$\sigma_\theta^2 = d \sum_{i=1}^{n} \lambda_i. \qquad (6)$$

The scattering density is defined as

$$\lambda(L_{rad}) \equiv \left(\frac{15}{p_0}\right)^2 \frac{1}{L_{rad}} \qquad (7)$$

where $p_0$ is the nominal momentum, chosen to be 3 GeV/c in this paper. For more information on these basics, refer to [7] and [15].

Let the reader note that Eqn. 3 and Eqn. 6 have the same form, i.e., the right side of these two equations is a linear integration of a parameter over the particle's path. Although muons are heavy charged particles, their trajectories through objects may be roughly treated as straight lines, even though multiple Coulomb scattering causes deviations. Thus, the scattering density $\lambda$ may be treated in a similar manner as the attenuation coefficient $\mu$ used in the computed tomography image reconstruction process.

Our setup in Geant4 is illustrated in Figure 3, with two pairs of detectors [25] vertically offset (by 100 cm) along the sides of a dry storage cask and a 10 cm separation between each pair of detectors. The detectors, each of dimension 350 cm wide by 150 cm high, are modeled as surfaces with perfect spatial and energy resolution. The simulated cask and associated spent fuel assemblies were configured using the design information of Sierra Nuclear's VSC-24 cask [26]. The "Muon Event Generator" was coupled with the Monte Carlo code Geant4[27]. In our implementation, the cask containing the spent fuel assemblies is fixed in location, while the detectors are allowed to rotate around the cask. In our simulation, we rotated the positions of the detectors at 2° increments to collect data from different views, or 90 times in total.

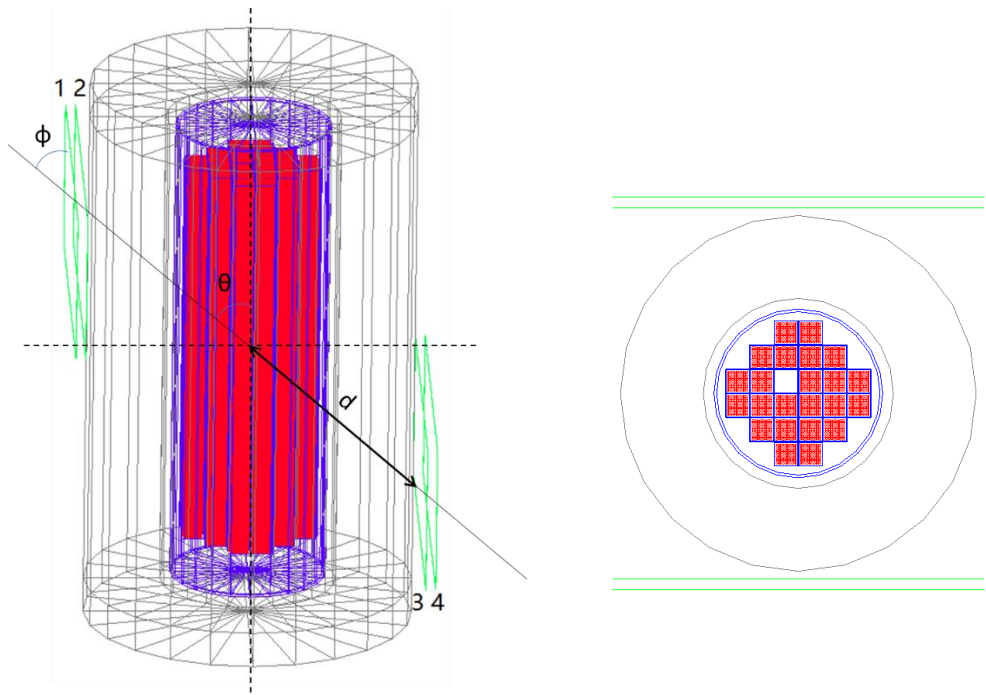

Figure 3. Side (left) and top-down (right) illustrations of the cask and detectors built in Geant4.

In our simulation, 4 detectors register the positions where each muon $j$ crosses. Let those positions be called $(x_{1j}, y_{1j}, z_{1j})$, $(x_{2j}, y_{2j}, z_{2j})$, $(x_{3j}, y_{3j}, z_{3j})$, and $(x_{4j}, y_{4j}, z_{4j})$ for the detectors shown in Figure 3, arranged from left to right. With four interaction points per muon, the absolute incident horizontal direction angles of each muon $\varphi_j$ can be calculated from Eqn (8), which is used to resort all registered muons into quasi-parallel ray data sets during data processing [28].

$$\varphi_j = angle((x_{2j} - x_{1j}), 1i * (y_{2j} - y_{1j})), \tag{8}$$

The scattering angles $\theta_j$ can be calculated using

$$\theta_j = acos(\frac{(x_{2j}-x_{1j}, y_{2j}-y_{1j}) \cdot (x_{4j}-x_{3j}, y_{4j}-y_{3j})}{|(x_{2j}-x_{1j}, y_{2j}-y_{1j})| \cdot |(x_{4j}-x_{3j}, y_{4j}-y_{3j})|}). \tag{9}$$

Using each muon's momentum to correct for the influence of polyenergetic muons and the recorded path length to correct for the influence of different trajectories [15], the normalized scattering angle of a muon becomes

$$\theta_j' = \frac{p_j}{p_0}\sqrt{\frac{D}{L_j}}\theta_j \tag{10}$$

where $p_0$ is the nominal momentum, $D$ is the vertical distance between detectors 2 and 3 (see Figure 3), and $L_j$ is the distance between $(x_{2j}, y_{2j}, z_{2j})$ and $(x_{3j}, y_{3j}, z_{3j})$.

Next, the registered incident muon spectrum is divided into one-degree-wide azimuthal bins according to their incident horizontal direction angles $\varphi$, which separates the incident muons into 180 quasi-parallel groups. For muons in each angular group, we project the muon scattering angles, according to each muon's incident horizontal direction and POCA point [29], to the plane that contains detector 3, as shown in Figure 4. Next, we divide this plane into 1 cm wide vertical bins along the horizontal direction and calculate the root mean square (RMS) of the muon scattering angles in each of these vertical bins to form a column of the sinogram. This differs from the method presented in [16], which only used the horizontal directions of the incident muon tracks to determine the position bin where the muon scattering angle is stored. Due to the fact that muon trajectory in an imaged object is not straight and the PoCA point roughly represents where the deflection of the muon is, we expect that projecting the scattering angle into the bin hit by the line passing by the PoCA point and along the incident horizontal direction (see Figure 4) should be more accurate.

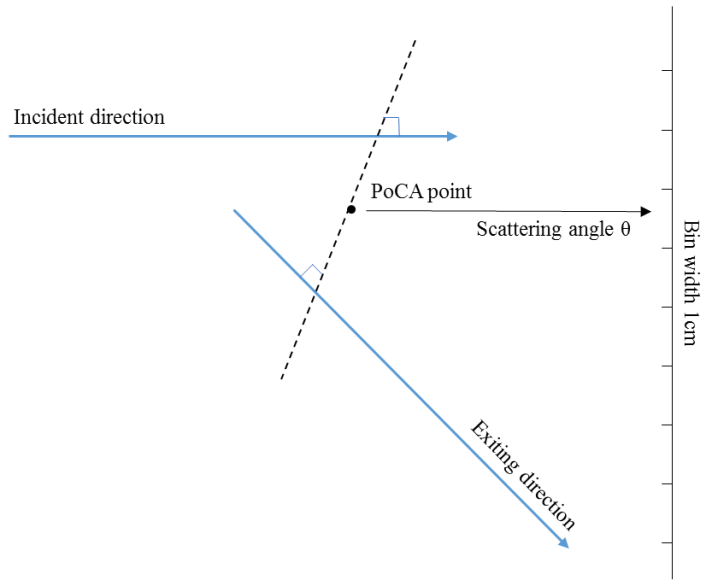

Figure 4. Illustration of a top-down view of incident and exiting trajectories, PoCA point, and the third detector plane showing the bin in which the scattering angle is stored.

## Muon CT reconstruction result analysis

For one of the middle four assemblies missing scenario, $7.1 \times 10^6$ muons are used for reconstruction, which is equivalent to 18.7 hours of exposure. For details of this calculation, refer to our calculation of muon collection time at the end of this paper. The root mean square of the scattering angles in each azimuthal bin is used to form the sinogram shown at left in Figure 5. Filtered back projection was used to reconstruct the image pictured at right in Figure 5, showing the estimated scattering density in each pixel.

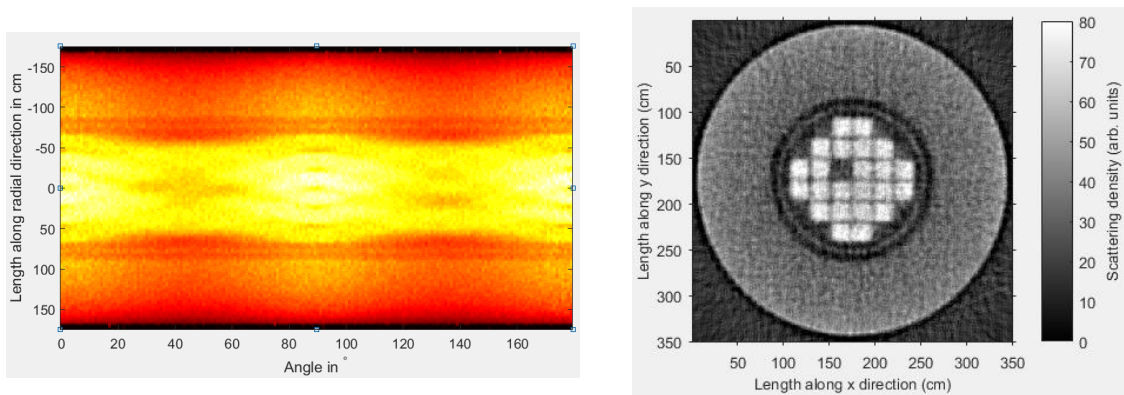

Figure 5. Sinogram (left) and reconstructed computed tomography image (right) of a dry storage cask with 1 fuel assembly missing (as in Figure 3).

Looking at the reconstructed image in Figure 5, it is evident that the middle fuel assembly is missing, which matches the configuration built in Geant4, as shown in

Figure *3*. The estimated scattering density for the spent nuclear fuel assembly is 68.0 ±2.7 arb. units, and the estimated value for the empty slot is 17.2±2.1 arb. units, which are separated by roughly 18.3 $\sigma$.

In order to be able to handle more challenging scenarios where one or more spent nuclear fuel assemblies from the center of cask are removed and replaced with dummy material in order to appear identical, we put forward some possible scenarios: (1) the middle four spent nuclear fuel assemblies are replaced by copper assemblies, or by lead assemblies, or by tungsten assemblies, or (2) one of the middle four spent nuclear fuel assemblies is replaced by a copper assembly, or by a lead assembly, or by a tungsten assembly.

In a case where the middle four spent nuclear assemblies are replaced, $7.5 \times 10^6$ muons are used for reconstruction. The corresponding Geant4 model and reconstructed images are shown in Figure *6*.

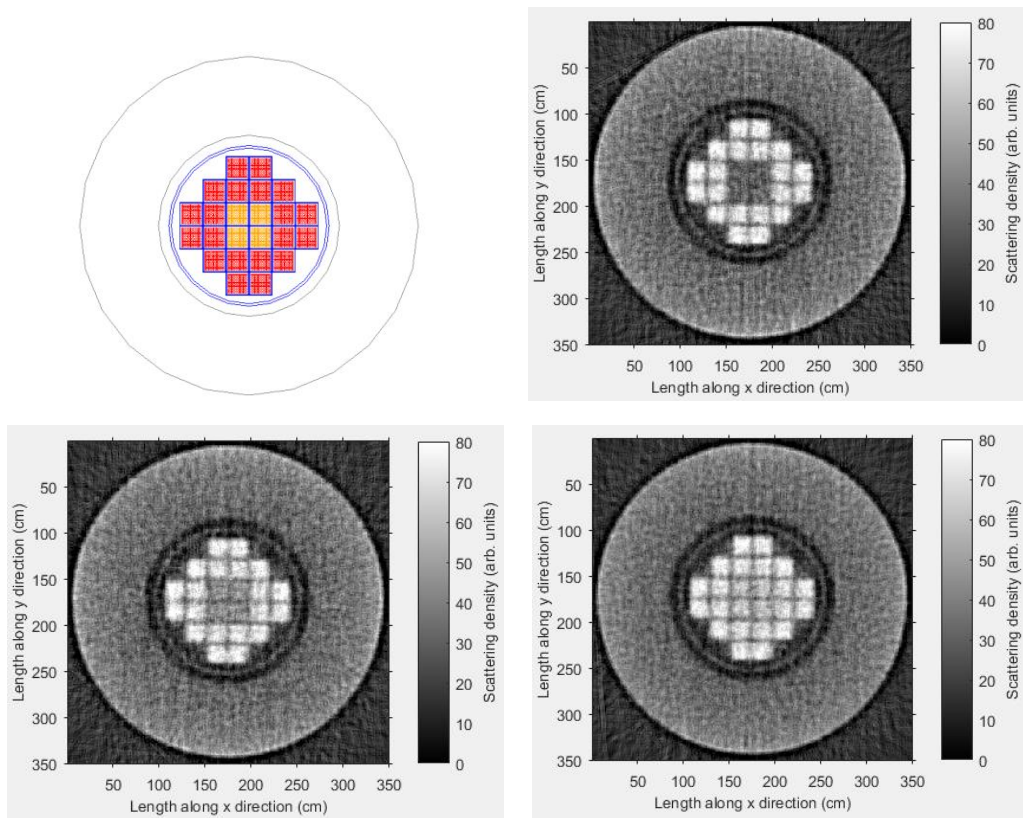

Figure 6. Geant4 model (upper left) and reconstructed images of a VSC-24 cask with the middle four spent fuel assemblies replaced by Cu (upper right), Pb (lower left) or W assemblies (lower right).

The estimated scattering densities of Fe (the main constituent of the canister), Cu, Pb, W and U (main constituent of spent nuclear fuel), are 25.8±2.1, 29.9±2.9,

45.7±2.4, 58.4±3.2 and 67.7±2.9 arb. units. In comparison, the known scattering densities (at muon momentum of 3 GeV/c) of these 5 materials are 14.2 (Fe), 17.4 (Cu), 71.3 (W), 44.5 (Pb), and 78.9 (U) mrad$^2$/cm. The relationship between known scattering densities and our estimated scattering densities is shown in Figure 7. A monotonically increasing relation between estimated and known scattering densities is expected. Yet, there is clearly some source of systematic error inherent to our estimation method that prevents us from estimating scattering density in an absolute sense. Even so, our results with an ideal detector system suggest there is potential to use muon imaging to find these scenarios to be "not as declared." Given the observed relationship, there is also potential to be able to identify the dummy material with some fidelity.

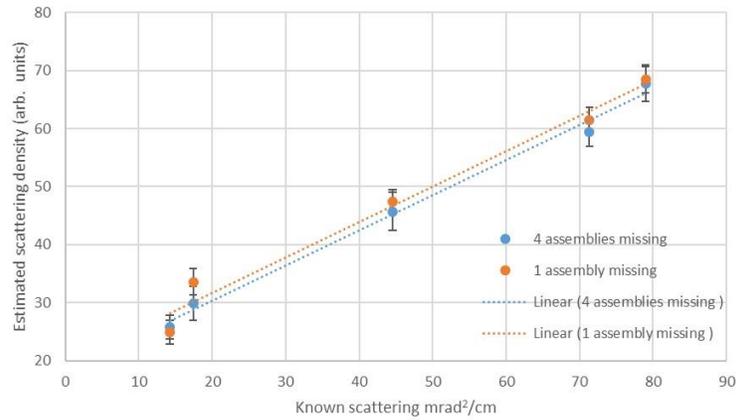

Figure 7. Comparison between known values and estimated values of scattering density for the scenarios shown in Figure 6 and Figure 8.

Furthermore, we aimed to understand the expected lower detection limit [30] by replacing only one of the middle four spent nuclear fuel assemblies with either a copper, lead, or tungsten assembly. For these scenarios, $10^7$ muons were used for reconstruction. The Geant4 model and reconstructed images are shown in Figure 8. The estimated scattering densities of Fe (the main constituent of the canister), Cu, Pb, W and U assemblies (the main constituent of spent nuclear fuel), are 25.1±2.0, 33.6 ±2.3, 47.4±2.1, 61.4±2.2 and 68.0±2.4 arb. units. Again, the known scattering densities of these 5 materials are 14.2 (Fe), 17.4 (Cu), 71.3 (W), 44.5 (Pb), and 78.9 (U) mrad$^2$/cm, so we are not currently able to determine the true scattering density values with our estimation method. Although we expect it to be more difficult to tell that there is one central assembly replaced by a dummy assembly than it is to tell that there are four replaced central assemblies, especially in the case of W, the statistical difference may be used to support the assertion that the spent nuclear fuel assembly is "not as declared." Also, it could be noticed that the estimated scattering density of the same materials in Figure 8 are bigger than that in Figure 6. This is due to the influence of surrounding materials and the inaccuracy of PoCA assumption. The more amount of high Z material surrounding a lower Z material, the latter is more likely to be overestimated. Because some of scattering points of surrounding high Z material fall in the region of low Z material.

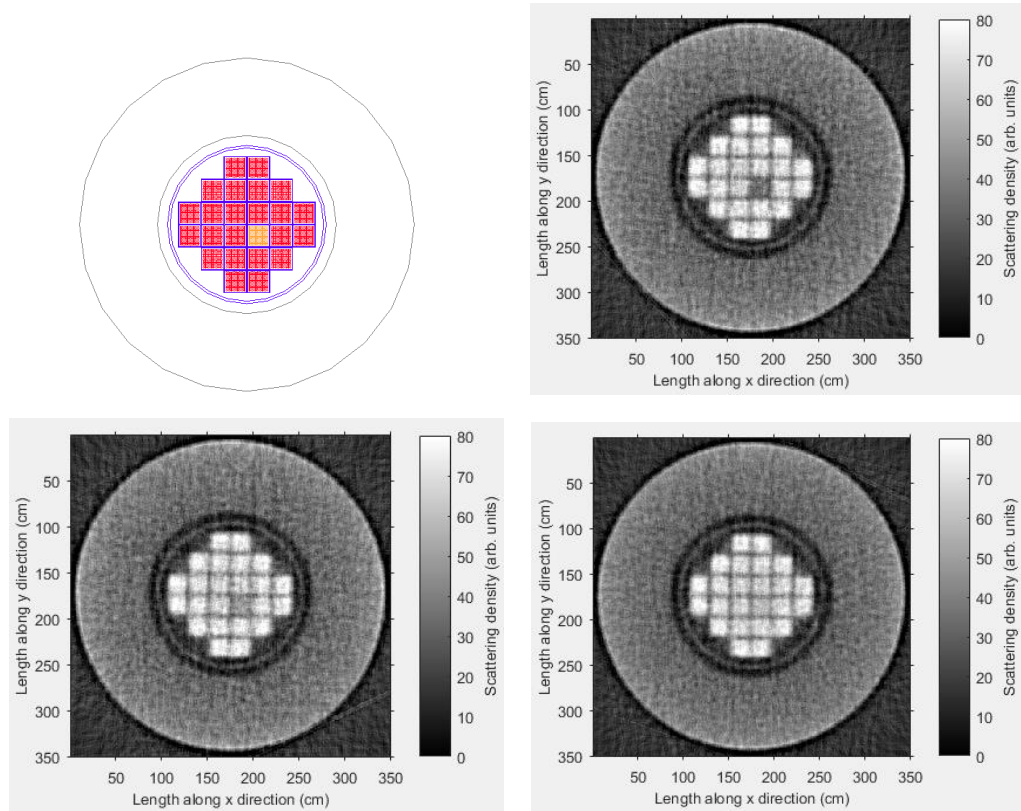

Figure 8. Geant4 model (upper left) and reconstructed images of a VSC-24 when one of middle assemblies is replaced by a Cu (upper right), Pb (lower left) or W (lower right) assembly.

## Calculation of muon collection time

In the configuration shown in Figure 3, both the zenith angle $\theta$ and detector angle $\Phi$ are 54.5 degree and distance $d$ is equal to 215.1 cm. When there is no simulated cask present between these two pairs of detectors, the muon flux rate registered by these detectors is $1.3 \times 10^4$ muons per minute. For the detailed steps used to calculate the flux rate, refer to [30]. About 2 GeV of energy will be lost by any muon that crosses our fully loaded dry storage cask [31]. Muons with initial momentum smaller than 2 GeV/c, accounting for about 30% of the total flux, tend to stop in the cask

[32]. Thus, the expected time needed to register $7.1\times10^6$ muons in our four detectors with the dry storage cask present is found to be 18.7 hours.

**Conclusions and future work**

In this paper, we describe how a computed tomography algorithm can be applied to image a VSC-24 dry storage cask using scattering angle as input information, instead of traditional transmission data, to yield geometry and estimate material information. Our calculation represents the limit of what is possible for the configuration described. Cosmic ray muons passing through the cask were observed by two pairs of ideal detectors vertically offset along the sides of the cask. When one of the middle four assemblies is removed, the reconstructed image is expected to clearly show the empty slot. We also showed that when the middle four assemblies were replaced by copper or lead or tungsten assemblies, a significant discrepancy is expected. Furthermore, when one of the middle four assemblies is replaced by a copper or lead or tungsten assembly, the estimated scattering densities are expected to be found to be "not as declared," because the dummy assemblies are expected to be separated from the surrounding spent nuclear assemblies by at least 3 standard deviations when $10^7$ muons are used. Since our method estimates the scattering density of any reconstructed pixel (see the relationship between estimated scattering densities and known scattering density presented in Figure 7), we also expect to be able to estimate the composition of the dummy material (see Figure 8).

**Acknowledgements**

This research was sponsored by the DOE Office of Nuclear Energy grant number DE-NE0008292. We would also like to sincerely thank Dr. Matt Durham of LANL, Dr. Chatzidakis Stylianos of ORNL and Alireza Rahimpour of the Department of Electrical Engineering and Computer Science at the University of Tennessee for helpful discussions.